\documentstyle[11pt]{article}

\newcommand{\beq}{\begin{equation}}
\newcommand{\eeq}{\end{equation}}
\newcommand{\T}{ $ T^{\mu\nu}_{;\nu} = 0 $ }
\newcommand{\p}{\partial}
\newcommand{\bfe}{\mbox{\bf e}}

\begin{document}
\vspace*{1.in}
\begin{center}

{\Large {\bf Einstein's Energy-Free Gravitational Field }} \\

\vspace{.5in}

{\large {\bf Kenneth Dalton }} \\
Post Office Box 587 \\
Mammoth Hot Springs \\
Yellowstone Park WY 82190, U.S.A. \\ 
\vspace{1.5in}

{\bf Abstract }

\end{center}
\vspace{.25in}

We show that Einstein's gravitational field has zero energy,
momentum, and stress.  This conclusion follows directly from
the gravitational field equations, in conjunction with the 
differential law of energy-momentum conservation
$ T^{\mu\nu}_{;\nu} = 0 $.  Einstein rejected this 
conservation law despite the fact that it is generally
covariant.  We trace his rejection to a misapplication
of Gauss' divergence formula.  Finally, we derive the
formula which pertains to energy-momentum conservation, viz.,
$ \oint {\bfe}_\mu \sqrt{-g}\, T^{\mu\nu} \, dV_\nu = 
\int {\bfe}_\mu\, T^{\mu\nu}_{;\nu} \sqrt{-g}\, d^4 x $.

\clearpage

\section*{\large {\bf 1. Einstein's rejection of the conservation
                     law \T }}
\indent

When he was faced with the equation

\beq
T^{\mu\nu}_{;\nu} =
      \frac{1}{\sqrt{-g}}
      \frac{\p \sqrt{-g}\,T^{\mu \nu}}{\p x^\nu}
      + \Gamma^\mu_{\nu\lambda} T^{\nu\lambda}
      = 0
\eeq
Einstein responded as follows:
the term $ \Gamma^\mu_{\nu \lambda} T^{\nu\lambda} $
``shows that laws of conservation of momentum and energy
do not apply in the strict sense for matter alone'' [1].
Quotations from the literature include Ref. [2]:
the equations \T\  ``are not what can properly be called
conservation laws''; Ref. [3]: the equation \T\ ``does not
generally express any conservation law whatever.''
These statements (and many others) constitute a very forceful
rejection of the generally covariant law.

\section*{\large {\bf 2. Proof of the conservation law \T }}

\indent

In the theory of special relativity, conservation
of energy-momentum is expressed by the Lorentz covariant
equation

\beq
   \frac{\p T^{\mu\nu} }{\p x^\nu} = 0
\eeq
Here, it is understood that flat rectangular coordinates
$ x^\mu = (x^0,x,y,z) $ are being used.  Suppose, instead,
that we choose ordinary flat spherical coordinates,
$ {x^\mu}' = (x^0,r,\theta,\phi) $.  What will be the law of
conservation in the new coordinate system?  To answer this
question, we begin with equation (2) and substitute the
transformed quantities

\beq
   T^{\mu\nu} =
      \frac{\p x^\mu}{\p {x^\alpha}'}
      \frac{\p x^\nu}{\p {x^\beta}'} {T^{\alpha\beta}}'
\eeq

\beq
 \frac{\p}{\p x^\nu} =
    \frac{\p {x^\alpha}'}{\p x^\nu} \frac{\p}{\p {x^\alpha}'}
\eeq
We then make use of

\beq
  {\Gamma^\mu_{\nu\lambda}}' =
     \frac{\p {x^\mu}'}{\p x^\alpha}
     \frac{\p x^\beta}{\p {x^\nu}'}
     \frac{\p x^\gamma}{\p {x^\lambda}'}
          \Gamma^\alpha_{\beta\gamma}
     + \frac{\p {x^\mu}'}{\p x^\alpha}
       \frac{\p^2 x^\alpha}{\p {x^\nu}' \p {x^\lambda}'}
\eeq

\beq
   \frac{1}{\sqrt{-g}\,'}
   \frac{\p \sqrt{-g}\,'}{\p {x^\lambda}'}
   = {\Gamma^\alpha_{\alpha\lambda}}'
\eeq
and arrive at the equation

\beq
       \frac{1}{\sqrt{-g}\,'}
       \frac{\p \sqrt{-g}\,'{T^{\mu\nu}}'}{\p {x^\nu}'}
       + {\Gamma^\mu_{\nu\lambda}}' {T^{\nu\lambda}}'
       = 0
\eeq
This proves the differential law of energy-momentum
conservation in the spherical coordinate system.
Because this law is generally covariant, it must hold true
for all systems of coordinates, flat or curved.

\section*{\large {\bf 3. The energy-free gravitational field}}

\indent

Einstein's gravitational field equations are

\beq
  R^{\mu\nu} - \frac{1}{2} g^{\mu\nu} R = \kappa\, T^{\mu\nu}
\eeq
$ T^{\mu\nu} $ is the stress-energy-momentum tensor of matter
and electromagnetism.
The covariant divergence of the left-hand
side is identically zero, therefore

\beq
   T^{\mu\nu}_{;\nu} = 0
\eeq
This equation means that the energy-momentum of matter and
electromagnetism is conserved, at all space-time points.
In other words, there is no
exchange of energy-momentum with the gravitational field.
We conclude that the gravitational field has no energy, 
momentum, or stress [4,5].

\section*{\large {\bf 4. Why was the covariant law rejected
                     in the past?}}
\indent

(a) In special relativity, the flow of energy-momentum is
described by means of Gauss' divergence formula

\beq
  \oint_R T^{\mu\nu} \, d^3 V_\nu
    = \int_R \frac{\p T^{\mu\nu}}{\p x^\nu} \, d^4 x
\eeq
The left-hand side is a closed surface integral over the
boundary of an infinitesimal region R; the right-hand side is
an integral over the interior of that region.
Energy-momentum is conserved in region R, if
$ {\p T^{\mu\nu}}/{\p x^\nu} = 0 $.

(b) In going over to curvilinear coordinates (flat or
curved), it was recognized that the surface elements
$ d^3 V_\nu $ must be multiplied by $ \sqrt{-g} $ in order to
form true vector components.  In this case, Gauss' theorem
yields

\beq
 \oint_R \sqrt{-g}\, T^{\mu\nu} \, d^3 V_\nu
 = \int_R \frac{\p \sqrt{-g}\, T^{\mu\nu}}{\p x^\nu} \, d^4 x
\eeq
This formula was used in the past to describe energy-
momentum flow.  It focussed attention upon the ordinary
divergence, $ {\p \sqrt{-g}\, T^{\mu\nu}}/{\p x^\nu} $,
and thus gave rise to a non-covariant law of conservation.
However, the expression $\sqrt{-g}\, T^{\mu\nu}\, d^3 V_\nu $
is a vector component with index $ \mu $, and its value
depends upon the arbitrary choice of coordinates.  Therefore,
it cannot fully represent any physical quantity; formula (11)
as it stands has no physical meaning.

(c) By introducing the system of basis vectors
$ {\bfe}_\mu $ , we obtain an expression which does not
depend upon the choice of coordinates:

\beq
 {\bfe}_\mu \sqrt{-g}\, T^{\mu\nu} \, d^3 V_\nu
 = {{\bfe}_\mu}' \sqrt{-g}\,' {T^{\mu\nu}}' \, d^3 {V_\nu}'
\eeq
This is the energy-momentum vector [6].  We now form the
vector sum 

\beq
\oint_R {\bfe}_\mu \sqrt{-g}\, T^{\mu\nu} \, d^3 V_\nu
  = \int_R \left\{
       {\bfe}_\mu \frac{\p \sqrt{-g}\, T^{\mu\nu}}{\p x^\nu}
  + ({\nabla}_\nu {\bfe}_\mu)
       \sqrt{-g}\, T^{\mu\nu} \right\} \, d^4 x
\eeq
over the infinitesimal region R.
\footnote{\small Equation (13) is not Gauss' formula.  
The left-hand side is a {\it vector} sum over the closed
boundary of region R.  The right-hand side involves the
rate of change of vectors $ {\bfe}_\mu $, as well as 
functions $ \sqrt{-g} \, T^{\mu\nu} $.  Gauss' formula
applies to functions, exclusively; it can therefore be
extended to cover a finite region.  Equation (13) is
confined to the infinitesimal region R; it yields a
differential law of conservation.}
The second term on the right is due to the basis vectors,
which change in magnitude and direction from point to point.
The rate of change is defined in terms of connection
coefficients $ \Gamma^\mu_{\nu\lambda} $  [6]

\beq
   {\nabla}_\nu {\bfe}_\mu
   = {\bfe}_\lambda \Gamma^\lambda_{\mu\nu}
\eeq
Substitute this expansion, then factor
$ {\bfe}_\mu $  and  $ \sqrt{-g} $ ,
in order to obtain the formula

\begin{eqnarray}
 \oint_R {\bfe}_\mu \sqrt{-g}\, T^{\mu\nu} \, d^3 V_\nu
 & = & \int_R {\bfe}_\mu \left\{ \frac{1}{\sqrt{-g}}
       \frac{\p \sqrt{-g}\, T^{\mu\nu}}{\p x^\nu}
   + \Gamma^\mu_{\nu\lambda} T^{\nu\lambda} \right\}
       \sqrt{-g} \, d^4 x  \nonumber \\
 & = & \int_R {\bfe}_\mu\, T^{\mu\nu}_{;\nu} \sqrt{-g}\, d^4 x
\end{eqnarray}
We conclude that energy-momentum is conserved in region R,
if \T\ [7,8].

\clearpage

\section*{\large {\bf 5. Summary}}

\begin{itemize}
\item[(a)] The differential law of conservation, \T , has
           been established:
    \begin{itemize}

    \item[(1)] by substituting flat curvilinear coordinates
         into the Lorentz covariant law,
    \item[(2)] by deriving the vector divergence formula for
         energy-momentum;
    \item[(3)] it is derived elsewhere by variation of the
         action under uniform displacement in space-time [7].

    \end{itemize}

\item[(b)] The principle consequence of this law is that
           Einstein's gravitational field has zero energy,
           momentum, and stress.

\item[(c)] The covariant law was rejected in the past.

\item[(d)] This rejection was based upon a physically
           meaningless application of Gauss' divergence
           formula.

\end{itemize}

\section*{\large {\bf References}}

\begin{enumerate}

\item A. Einstein, ``The Foundation of the General Theory of
      Relativity'' in {\it The Principle of Relativity}
      (Dover, New York, 1952) section 18.
\item P. Bergmann, {\it Introduction to the Theory of
      Relativity} (Dover, New York, 1976) page 194.
\item L. Landau and E. Lifshitz, {\it The Classical Theory of
      Fields} (Pergamon, Elmsford, 4th ed., 1975) section 96.
\item K. Dalton, {\it Hadronic J.} {\bf 17}, 139 (1994).
\item K. Dalton, http://publish.aps.org/eprint/aps1997feb28 004/
\item C. Misner, K. Thorne, and J. Wheeler, {\it Gravitation}
      (Freeman, New York, 1973) pages 151, 259.
\item K. Dalton, {\it Gen.Rel.Grav.} {\bf 21}, 533 (1989).
\item J. Vargas and D. Torr, {\it Gen.Rel.Grav.} {\bf 23},
      713 (1991).

\end{enumerate}

\end{document}